\documentclass[aps,twocolumn,prd,dvipsnames,usenames,floatfix,superscriptaddress,nofootinbib,longbibliography,preprintnumbers]{revtex4-1}

\usepackage{graphicx}
\usepackage{amsmath}
\usepackage{amsfonts}
\usepackage[colorlinks=true,linktocpage=true,urlcolor=blue,citecolor=blue,linkcolor=blue]{hyperref}

\usepackage{amsmath}
\usepackage{amsfonts}

\usepackage{tabularx}
\newcolumntype{C}{>{\centering\arraybackslash}X}
\newcolumntype{R}{>{\raggedleft\arraybackslash}X}

\usepackage{dsfont}

\usepackage[usenames, dvipsnames]{color}
\usepackage[normalem]{ulem}
\usepackage{xcolor}

\newcommand{\be}{\begin{eqnarray}}

\newcommand{\ee}{\end{eqnarray}}

\definecolor{colorA}{HTML}{1E90FF}
\definecolor{colorB}{HTML}{228B22}
\definecolor{colorC}{HTML}{FF7F00}
\definecolor{colorD}{HTML}{4B0082}
\definecolor{colorE}{HTML}{B22222}

\definecolor{lgreen}{HTML}{32CD32}
\definecolor{lgray}{HTML}{D3D3D3}
\definecolor{dblue}{HTML}{1E90FF}
\definecolor{dblue}{HTML}{1E90FF}
\definecolor{orange}{HTML}{FF4500}
\definecolor{indigo}{HTML}{4B0082}

\definecolor{teal}{HTML}{008080}
\definecolor{firebrick}{HTML}{B22222}
\definecolor{salmon}{HTML}{FA8072}
\definecolor{darkgreen}{HTML}{006400}

\newcommand{\jhu}{William H. Miller III Department of Physics and Astronomy, Johns Hopkins University, Baltimore, MD 21218, USA}
\newcommand{\texas}{Weinberg Institute for Theoretical Physics, The University of Texas at Austin, Austin, TX 78712, USA}
\newcommand{\cornell}{Laboratory for Elementary Particle Physics Cornell University, Ithaca, NY 14853, USA}
\newcommand{\korea}{Department of Physics, Korea University Seoul 02841, Republic of Korea}

\graphicspath{ {plots/} }

\makeatletter
\def\doauthor#1#2#3{%
  \ignorespaces#1\unskip
  \begingroup
   #3%
  \@if@empty{#2}{\@listcomma\endgroup{}{}}{\endgroup{\comma@space}{}\frontmatter@footnote{#2}}%
  \space \@listand
}%
\makeatother

\makeatletter
\def\@ssect@ltx#1#2#3#4#5#6[#7]#8{%
  \def\H@svsec{\phantomsection}%
  \@tempskipa #5\relax
  \@ifdim{\@tempskipa>\z@}{%
    \begingroup
      \interlinepenalty \@M
      #6{%
       \@ifundefined{@hangfroms@#1}{\@hang@froms}{\csname @hangfroms@#1\endcsname}%
       {\hskip#3\relax\H@svsec}{#8}%
      }%
      \@@par
    \endgroup
    \@ifundefined{#1smark}{\@gobble}{\csname #1smark\endcsname}{#7}%
  }{%
    \def\@svsechd{%
      #6{%
       \@ifundefined{@runin@tos@#1}{\@runin@tos}{\csname @runin@tos@#1\endcsname}%
       {\hskip#3\relax\H@svsec}{#8}%
      }%
      \@ifundefined{#1smark}{\@gobble}{\csname #1smark\endcsname}{#7}%
      \addcontentsline{toc}{#1}{\protect\numberline{}#8}%
    }%
  }%
  \@xsect{#5}%
}%
\makeatother

\begin{document}

\title{\vspace{-0.2cm}A Generative Modeling Approach to Reconstructing 21-cm Tomographic Data}

\author{Nashwan Sabti$^{\mathds{S},}$}
\affiliation{\jhu}
\author{Ram Reddy$^{\mathds{R},}$}
\affiliation{\texas}
\author{Julian B. Mu\~noz$^{\mathds{M},}$}
\affiliation{Department of Astronomy, The University of Texas at Austin, 2515 Speedway, Stop C1400, Austin, TX 78712, USA}
\author{Siddharth Mishra-Sharma$^{\mathds{MS},}$}
\affiliation{The NSF AI Institute for Artificial Intelligence and Fundamental Interactions}
\affiliation{Center for Theoretical Physics, Massachusetts Institute of Technology, Cambridge, MA 02139, USA}
\affiliation{Department of Physics, Harvard University, Cambridge, MA 02138, USA}
\author{Taewook Youn$^{\mathds{Y},}$}
\affiliation{\cornell}
\affiliation{\korea}

\def\thefootnote{$\mathds{S}$\hspace{1.9pt}}\footnotetext{\href{mailto:nash.sabti@gmail.com}{nash.sabti@gmail.com}}
\def\thefootnote{$\mathds{R}$\hspace{1.2pt}}\footnotetext{\href{mailto:ramreddy@utexas.edu}{ramreddy@utexas.edu}}
\def\thefootnote{$\mathds{M}$\hspace{1.2pt}}\footnotetext{\href{mailto:julianbmunoz@utexas.edu}{julianbmunoz@utexas.edu}}
\def\thefootnote{$\mathds{MS}$\hspace{-0.35pt}}\footnotetext{\href{mailto:smsharma@mit.edu}{smsharma@mit.edu}}
\def\thefootnote{$\mathds{Y}$\hspace{1.6pt}}\footnotetext{\href{mailto:taewook.youn@cornell.edu}{taewook.youn@cornell.edu}}

\setcounter{footnote}{0}
\def\thefootnote{\arabic{footnote}}

\begin{abstract}
\noindent Analyses of the cosmic 21-cm signal are hampered by astrophysical foregrounds that are far stronger than the signal itself. These foregrounds, typically confined to a wedge-shaped region in Fourier space, often necessitate the removal of a vast majority of modes, thereby degrading the quality of the data anisotropically. To address this challenge, we introduce a novel deep generative model based on stochastic interpolants to reconstruct the 21-cm data lost to wedge filtering. Our method leverages the non-Gaussian nature of the 21-cm signal to effectively map wedge-filtered 3D lightcones to samples from the conditional distribution of wedge-recovered lightcones. We demonstrate how our method is able to restore spatial information effectively, considering both varying cosmological initial conditions and astrophysics. Furthermore, we discuss a number of future avenues where this approach could be applied in analyses of the 21-cm signal, potentially offering new opportunities to improve our understanding of the Universe during the epochs of cosmic dawn and reionization.  
\\\\
{\centering\noindent \href{https://github.com/NNSSA/Rec21}{\raisebox{-1pt}{\includegraphics[width=9pt]{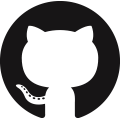}}}\hspace{2pt} Code, pre-trained models, and scripts for making plots in this paper can be found \href{https://github.com/NNSSA/Rec21}{here}.}
\end{abstract}

\maketitle

\section{Introduction}
\label{sec:introduction}

The high-redshift 21-cm signal serves as a promising tool for investigating the early Universe during the epochs of cosmic dawn and reionization~\cite{Furlanetto:2006jb,Morales_2010,Pritchard:2011xb,Liu:2019awk}. Unlike direct observations of galaxies, 21-cm experiments map the distribution of neutral hydrogen throughout vast volumes of the Universe. Current radio interferometers such as HERA~\cite{DeBoer:2016tnn} are on the cusp of achieving a statistical detection of the 21-cm signal by analyzing its power spectrum~\cite{HERA:2021bsv, HERA:2022wmy}. The ultimate goal of these studies, however, is to advance toward tomographic 3D imaging of its spatial fluctuations~\cite{Furlanetto:2004zw, Koopmans:2015sua, Mellema:2015zha}, which would provide significant insights into the early Universe's structure and evolution.

Detecting the 21-cm signal is significantly hindered by the presence of astrophysical foregrounds that are several orders of magnitude stronger than the signal itself, see e.g. Refs.~\cite{Santos:2004ju, Bowman:2008mk}. These foregrounds, primarily sourced from galactic and extra-galactic synchrotron and free-free emissions, tend to be spectrally smooth~\cite{Oh:2003jy}. As a result, in Fourier space, they mainly contaminate the long-wavelength modes perpendicular to the line of sight, making their separation in theory a relatively straightforward task. In practice, however, an interferometer's frequency-dependent response to such emissions introduces an additional, non-smooth component in Fourier space~\cite{Morales_2012, Parsons_2012, Vedantham_2012,Pober:2013ig,Hazelton:2013xu,Thyagarajan:2013eka,Liu:2014bba,Liu:2014yxa}. This anisotropic component occupies a wedge-shaped region and, consequently, only a small window in Fourier space remains where the 21-cm signal can be observed without substantial foreground contamination. 

While completely removing the foreground-dominated regions in Fourier space is always possible, this would clearly have a detrimental effect on the signal-to-noise of the information extracted with this data (see e.g. Ref.~\cite{Seo:2015aza}). Moreover, if we are aiming for 3D tomographic imaging of the 21-cm signal, it is essential to retain as many Fourier modes as possible. This is because wedge filtering removes Fourier modes in an anisotropic way, leading to a distortion of the original image beyond simple blurring and making an interpretation of what we see even more complicated~\cite{Beardsley:2014bea}.

\begin{figure*}[ht!]
    \centering
    \includegraphics[width=\linewidth]{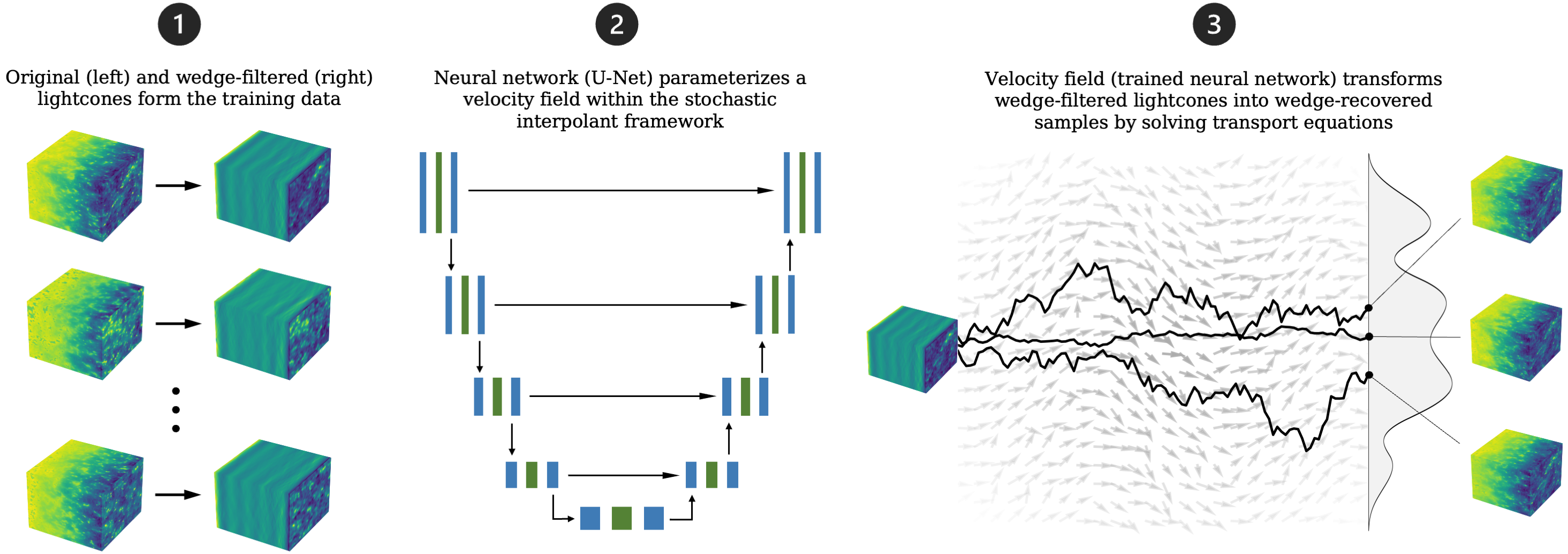}
    \caption{A schematic overview of our approach to develop a generative model aimed at reconstructing tomographic 21-cm maps at the field level from wedge-removed data. This method encompasses several stages, including data generation, neural-network training, and solving transport equations. \textbf{{Part 1:}} We generate 21-cm lightcones using simulations and remove Fourier modes that are contaminated by foregrounds, resulting in pairs of `original' and wedge-filtered lightcones. \textbf{Part 2:} These data pairs are then used within the stochastic interpolant framework, where a neural network is trained to learn a time-dependent velocity field that can dynamically transport a wedge-filtered lightcone to a wedge-recovered lightcone. \textbf{Part 3:} A stochastic differential equation is solved, with the trained neural network acting as a drift function, to push an input wedge-filtered lightcone to a sample from the distribution of wedge-recovered lightcones conditioned on this input.} 
    \label{fig:sketch}
\end{figure*}

To address the challenge posed by foreground contamination, various strategies beyond avoidance have been proposed. These include forward modeling the effect of the wedge within a probabilistic framework, e.g.~\cite{Kern2023, Saxena:2024rhu}, and accurately modeling the noise for the purpose of subtraction~\cite{Shaw:2013wza, Kern:2020kky, Shaw:2014khi} (although a challenging task, see Refs.~\cite{Liu:2019awk,Pober:2013jna} for a summary). More recently, a promising set of approaches based on machine learning has emerged~\cite{Gillet:2018fgb, Hassan:2019cal, Li:2019znt,Makinen:2020gvh,Villanueva-Domingo:2020wpt,Prelogovic:2021ljp,Gagnon-Hartman:2021erd, Kennedy:2023zos, Shi:2023nrk, Bianco:2023eec, Mertens:2023dcl}. One such approach leverages neural networks to identify correlations between Fourier modes inside and outside of the wedge region~\cite{Gagnon-Hartman:2021erd}, which is largely possible due to the inherent non-Gaussian nature of the 21-cm signal~\cite{Gorce:2019tkj, Shimabukuro:2015iqa,Majumdar:2017tdm,Watkinson:2018efd,Hutter:2019yta}. Then, under certain conditions, given a wedge-filtered image it becomes feasible for the network to reconstruct a significant portion of the modes that were originally lost in the filtering process. A major advantage of this approach is that it starts with relatively uncontaminated 21-cm data and does not require precise noise modeling.

In this paper, we build upon the work of Ref.~\cite{Gagnon-Hartman:2021erd} by exploring a deep-learning method based on generative modeling to reconstruct the modes inside the wedge region. Ref.~\cite{Gagnon-Hartman:2021erd} approached the problem as a classification task, where the main goal was to recover a binarized map of the 21-cm signal that indicates which regions are neutral and ionized. By construction, this method is deterministic: each input wedge-filtered box produces a unique wedge-recovered output box. In contrast, here we draw on recent advances in machine learning in building new classes of generative models that unify diffusion- and flow-based methods~\cite{albergo2023stochasticinterpolantsunifyingframework, albergo2023stochasticinterpolantsdatadependentcouplings}. Specifically, we make use of stochastic processes known as `stochastic interpolants' that are able to continuously bridge two arbitrary probability densities\footnote{Variants of the stochastic interpolant method are also referred to as rectified flows~\cite{liu2022flowstraightfastlearning} and flow matching~\cite{lipman2023flowmatchinggenerativemodeling}.}. Within our context, this method makes it possible to connect the distribution of wedge-filtered 21-cm data to that of wedge-recovered data. The stochastic nature of this process allows a fixed input (i.e., the same wedge-filtered 21-cm lightcone) to be transformed into a distribution of possible wedge-recovered lightcones that is conditioned on the input. In this way, we can sample from this distribution and obtain realizations of wedge-recovered solutions --- a generative model. 

We summarize the main differences between our approach and previous works: \emph{(1)} Our method utilizes a generative model to obtain samples from the conditional distribution of wedge-recovered lightcones, \emph{(2)} we do not consider binarized maps, but rather allow for the recovery of the full 21-cm brightness temperature, \emph{(3)} both cosmological initial conditions and a range of astrophysical parameters are varied in the data generation process, which results in the final distribution being marginalized over these effects, and \emph{(4)} lightcones are considered instead of coeval boxes (where all voxels share the same redshift), as they more closely resemble real data and enable recovery over a range of redshifts. We provide an illustrative overview of our procedure in Fig.~\ref{fig:sketch}. 

Through this approach, we aim to provide a more robust and flexible framework for reconstructing 21-cm maps. For instance, dealing with distributions enables us to propagate uncertainties in the recovered lightcones into downstream tasks, which, as we will show, can be particularly useful for deriving the power spectrum of the 21-cm signal. This may have important implications for the measurement accuracy of astrophysical or cosmological quantities of significant interest, such as the Hubble expansion rate at high redshifts. Additionally, this method could potentially open up avenues for extracting new information from the data by combining it with other types of observations. For example, a reconstructed 21-cm tomographic map allows for improved cross-correlations with other datasets, including photometric galaxy surveys~\cite{LaPlante:2022nlp, Hutter:2023rja}, Lyman-$\alpha$ forest observations~\cite{Carucci:2016yzq}, and CMB anistropy and lensing maps~\cite{Tashiro_2010, Tanaka:2019nph}.

\begin{figure}[t!]
    \centering
    \includegraphics[width=\linewidth]{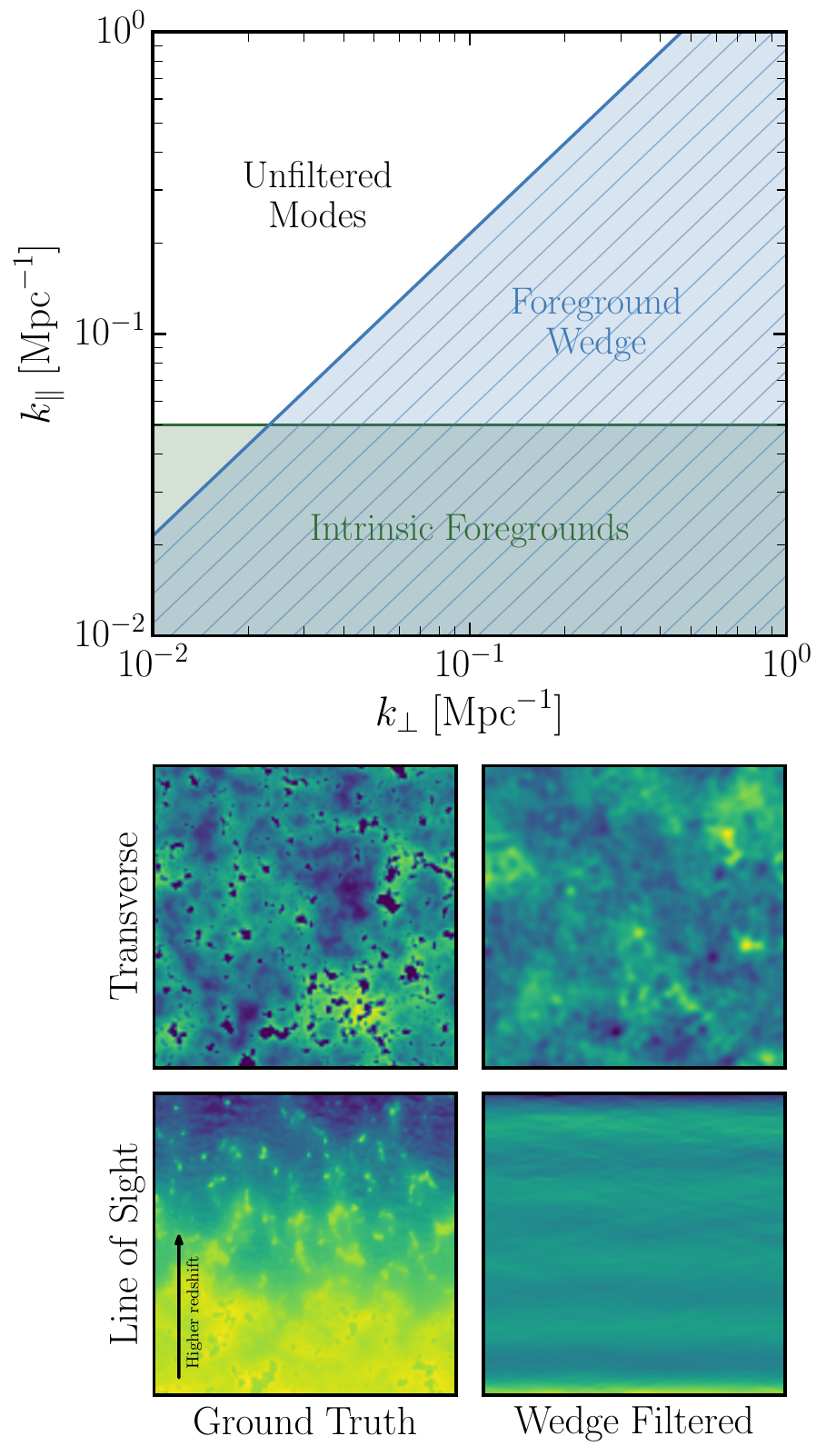}
    \caption{\textbf{Top:} Illustration of the Fourier modes affected by foreground contamination considered in this work. Astrophysical foregrounds occupy the low $k_{\parallel}$ modes (green), while the instrument response causes leakage to higher $k_{\parallel}$ modes (blue, hatched). \textbf{Bottom:} Sample slices of ground-truth and wedge-filtered lightcones along the transverse (redshift $\approx 9.1$) and line-of-sight directions.}
    \label{fig:wedge_plots}
\end{figure}

This paper is organized as follows: In Sec.~\ref{sec:methodology}, we detail our methodology, covering dataset generation, the wedge-removal process, the application of stochastic interpolants, an overview of our neural network architecture, our training and evaluation procedures, culminating with a description of the generation of reconstructed samples. In Sec.~\ref{sec:results}, we present our results, which include a demonstration of the wedge reconstruction process in image space as well as in Fourier space through the power spectrum. In Sec.~\ref{sec:future}, we discuss current limitations of our method, potential improvements, and explore several relevant applications. Finally, we conclude in Sec.~\ref{sec:conclusion}.

Throughout the paper, we will fix cosmological parameters to the Planck 2018 results~\cite{Planck:2018vyg}: $h = 0.6727$, $\sigma_8 = 0.8120$, $\Omega_\mathrm{m} = 0.3166$, $\Omega_\mathrm{b} = 0.0494$, and $n_\mathrm{s} = 0.9649$.

\section{Methodology}
\label{sec:methodology}

\subsection{Dataset and Forward Model}
\label{subsec:data_and_model}
We make use of the code \texttt{21cmFAST}~\cite{Mesinger:2010ne,Murray:2020trn} to generate lightcones of the 21-cm brightness temperature. These lightcones have a volume of $(512)^3 \,\mathrm{Mpc}^3$ and feature a resolution of $4\,\mathrm{Mpc}$ along each axis, resulting in 128 voxels per axis. We post-process each box such that the line-of-sight axis has a redshift range $z \in [8.9, 11.3]$ and is centered at $z \approx 10$. We keep the default settings of \texttt{21cmFAST}, modifying only cosmological initial conditions and a few key astrophysical parameters for each simulation. Specifically, we vary the following parameters within logflat priors whose limits are informed by the wedge-removed power-spectrum forecast done in Ref.~\cite{Mason:2022obt}: The fraction of galactic gas in stars $\log_{10}f_{\star,10} \in [-1.45, -1.15]$, the escape fraction of ionizing photons $\log_{10}f_{\mathrm{esc},10} \in [-1.17, -0.84]$, and the specific X-ray luminosity $\log_{10}l_X \in [40.45, 40.55]$. Although more parameters could be varied and broader prior ranges could be considered, such an extension would require a larger training set and more computational resources, which we leave for future work. In its current implementation, our approach likewise does not incorporate Population III stars (i.e., mini halos), but they can be straightforwardly included following the methodology outlined here.

We generate a total of 25,000 lightcones, each featuring unique cosmological initial conditions and astrophysical parameters as previously mentioned. Since redshift-space distortions are included in the simulated boxes, we augment them by rotating the boxes four times by $90^\circ$ only along the line-of-sight axis, obtaining a total of 100,000 lightcones. Such a large number of lightcones is eventually necessary for the neural network to effectively learn correlations and be able to reconstruct modes starting from an unseen wedge-filtered lightcone.

We then process the lightcones by removing modes dominated by intrinsic foregrounds and those within the foreground wedge, see top panel in Fig.~\ref{fig:wedge_plots}. We employ the notation $k_\parallel \equiv k_z$ and $k_\perp \equiv \sqrt{k_x^2 + k_y^2}$ to represent the Fourier modes along the line-of-sight and those perpendicular to it, respectively. These can be thought of as the Fourier analogs of redshift (or frequency) and the separation angle. To address intrinsic foregrounds, we remove all modes with $|k_\parallel| < 0.05\,\mathrm{Mpc}^{-1}$~\cite{Pober:2013ig, Kennedy:2023zos}. We define the wedge region as $|k_\parallel| < |k_\perp|\tan\psi$, where $\psi$ is the wedge angle. The efficacy of the reconstruction is affected by the value of $\psi$; for example, a smaller $\psi$ means that fewer modes are nullified and, consequently, it would be easier to reconstruct the original lightcone. Since the exact value of $\psi$ is experiment- and configuration-specific, in the remainder of the paper we will fix it to $\psi = 65^\circ$ for demonstrative purposes\footnote{This means that we discard roughly 80\% of all modes in Fourier space.} and refer to Ref.~\cite{Gagnon-Hartman:2021erd} for a more detailed discussion. Throughout the text, we will refer the removal of both types of foregrounds as `wedge filtering'. After nullifying the relevant modes, we apply the inverse Fourier transform to obtain the wedge-filtered lightcones back in image space. We show the impact of the removal in the bottom plots of Fig.~\ref{fig:wedge_plots}. The top and bottom rows show slices perpendicular to and along the line-of-sight axis. The left columns are before wedge filtering; the right columns are after. With the way the wedge is defined, features on small scales experience stronger filtering, while modes along the line of sight are subjected to a more intense low-cut filter. We find in both cases that only a distorted version of the original lightcone is left intact. This makes the interpretation of imaging data a complicated task.

\subsection{Stochastic Interpolants}
\label{subsec:stochastic_interpolants}
Our dataset consists of pairs of 21-cm lightcones and the corresponding wedge-filtered lightcones, defining the joint density $(x_{21}, x_{21}^{\mathrm{wf}}) \sim \rho(x_{21}, x_{21}^{\mathrm{wf}})$. These are our target and base quantities, respectively, represented as voxelized grids described in the previous section. The stochastic interpolant framework~\cite{albergo2023buildingnormalizingflowsstochastic,albergo2023stochasticinterpolantsdatadependentcouplings} provides a way to build generative models based on diffusions or flows to dynamically transport a base measure ($x_{21}^{\mathrm{wf}}$) onto a target ($x_{21}$). The goal of the generative model is to sample from the conditional probability density $\rho(x_{21}|x_{21}^\mathrm{wf})$ of $x_{21}$ given $x_{21}^{\mathrm{wf}}$. We do this by solving a stochastic differential equation (SDE) that maps a fixed initial condition, $X(t=0) = x_{21}^\mathrm{wf}$, to a sample from the conditional distribution, $X(t=1)\sim \rho(x_{21}|x_{21}^\mathrm{wf})$, where $t \in [0,1]$ is a time variable. The time $t$ has no physical meaning in the current context, and simply parameterizes the intermediate marginal densities. Repeating this procedure many times then produces an ensemble of samples that describe the distribution of wedge-recovered lightcones consistent with the wedge-filtered input. Note that unlike vanilla diffusion models~\cite{sohldickstein2015deepunsupervisedlearningusing,ho2020denoisingdiffusionprobabilisticmodels,song2020generativemodelingestimatinggradients,song2021scorebasedgenerativemodelingstochastic, yang2024diffusionmodelscomprehensivesurvey}, which would require conditioning on the wedge-filtered lightcone during the reverse process due to their Gaussian base distribution, our approach eliminates the need for such conditioning.

We construct the stochastic interpolant $x(t)$ that bridges two probability densities as follows~\cite{chen2024probabilisticforecastingstochasticinterpolants}:
\begin{align}
    \label{eq:stochastic_interpolant}
    x(t) = \alpha(t)x_{21}^\mathrm{wf} + \beta(t)x_{21} + \sigma(t)W(t)\ ,
\end{align}
where $\alpha(t) = \sigma(t) = 1 - t$, $\beta(t) = t^2$, and $W$ is a Wiener process. Note that $\alpha(t)$, $\beta(t)$, and $\sigma(t)$ satisfy boundary conditions such that $x(0) = x_{21}^\mathrm{wf}$ and $x(1) = x_{21}$. The intermediate time steps include a random element due to the random process $W$. Next, it can be shown that the velocity field corresponding to the interpolant in Eq.~\eqref{eq:stochastic_interpolant} is given by~\cite{chen2024probabilisticforecastingstochasticinterpolants}:
\begin{align}
    \label{eq:drift_function}
    v(t, x_{21}, x_{21}^\mathrm{wf}) = \dot{\alpha}(t)x_{21}^\mathrm{wf} + \dot{\beta}(t)x_{21} + \dot{\sigma}(t)W(t)\ ,
\end{align}
with the dot denoting derivation with respect to time $t$. We approximate this function with a neural network $\hat{v}(t, x(t))$ by minimizing the following objective with stochastic gradient descent~\cite{chen2024probabilisticforecastingstochasticinterpolants}:
\begin{align}
    \label{eq:loss}
    L[\hat{v}] = \int_0^1 \mathrm{d}t\,  \mathbb{E}\left[\left\vert\hat{v}(t, x(t)) - v(t, x_{21}, x_{21}^\mathrm{wf})\right\vert^2\right]\ ,
\end{align}
where the expectation value is taken over the data pairs $(x_{21}, x_{21}^{\mathrm{wf}}) \sim \rho(x_{21}, x_{21}^{\mathrm{wf}})$ and $W$. The time integral can be simply evaluated through Monte Carlo sampling. The learned velocity field will then act as a drift function in the stochastic differential equation~\cite{song2021scorebasedgenerativemodelingstochastic}, 
\begin{align}
    \label{eq:sde}
    \mathrm{d}X(t) = \hat{v}(t, X(t))\mathrm{d}t + \sigma(t)\mathrm{d}W(t)\ ,
\end{align}
whose solutions are such that $X(t=1) \sim \rho(x_{21}|x_{21}^\mathrm{wf})$. Here, $W(t)$ is another Wiener process. In other words, our neural network represents the velocity field that pushes a wedge-filtered lightcone to a sample from the distribution of wedge-recovered lightcones (which itself is an approximation of the `true' distribution of $x_{21}$ conditioned on $x_{21}^\mathrm{wf}$). The dynamics are governed by the SDE in Eq.~\eqref{eq:sde}, which we will further discuss in Sec.~\ref{subsec:training_evaluation}.

\subsection{Neural Network Architecture}
\label{subsec:architecture}
Given the intrinsic non-Gaussian nature of the 21-cm signal, our neural network architecture should be able to effectively learn correlations in the 21-cm lightcones to reconstruct modes within the wedge region. This is akin to super-resolution problems, where the goal is to create a high-resolution image from a low-resolution version~\cite{Yang_2019}. Convolutional neural networks (CNNs) are appropriate for such tasks, see e.g. Refs.~\cite{shi2016realtimesingleimagevideo,Dong2016}, and they will thus serve as the backbone of our architecture, with the primary objective to represent the velocity field in Eq.~\eqref{eq:drift_function} by capturing the non-Gaussian correlations.

First, we take a data pair $(x_{21}, x_{21}^\mathrm{wf})$ and assign a random time $t$ sampled from a uniform distribution $\mathcal{U}(0,1)$ to construct the stochastic interpolant $x(t)$ as in Eq.~\eqref{eq:stochastic_interpolant}. This sampling ensures that the model covers the whole time range of interest during training. These two quantities, $t$ and $x(t)$, are the inputs to our model. We project the time variable $t$ onto an embedding space of dimension 48 using sinusoidal functions~\cite{vaswani2023attentionneed}, and then process it through a multi-layer perceptron consisting of two linear layers of hidden dimension 192 with SiLU activation functions. To aid in capturing temporal dependencies, the model is conditioned on this time variable by adding a linear projection of the embedded $t$ to the output of a convolutional block as described below.

\begin{figure*}[ht!]
    \centering
    \includegraphics[width=\textwidth]{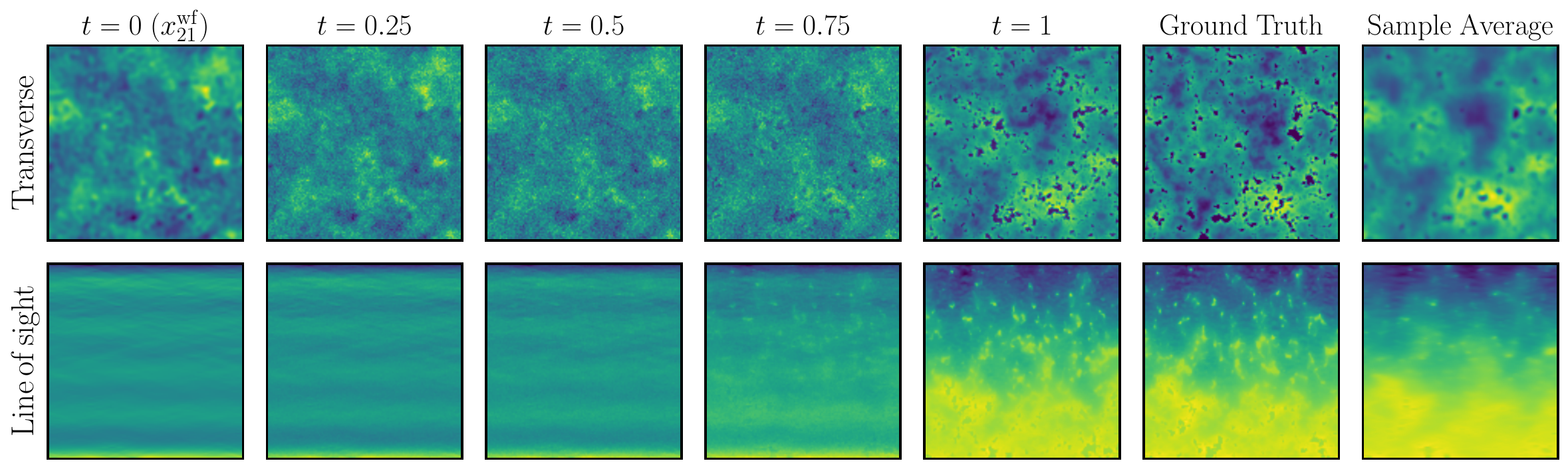}
    \caption{Reconstruction of a wedge-filtered lightcone over time. Top and bottom rows show a slice perpendicular to ($z\approx 9.1$) and along the line-of-sight axis, respectively. The leftmost column is a wedge-filtered input ($x_{21}^\mathrm{wf}$) at time $t=0$. Over time, this input gets transported to a sample from the conditional distribution $\rho(x_{21}|x_{21}^\mathrm{wf})$ at $t=1$, which can be compared to the ground truth in the sixth column. The last column shows the average of 2000 sampled lightcones.}
    \label{fig:reconstruction}
\end{figure*}

Then, we put the stochastic interpolant through the main structure of our model, which is a 3D U-Net neural network~\cite{ronneberger2015unetconvolutionalnetworksbiomedical}. The U-Net consists of 4 encoding blocks, followed by a bottom block, and finally 4 decoder blocks. Each of these blocks is made out of two pre-activation `convolutional blocks', which is a sequence of an instance normalization layer, a ReLU activation, and a 3D convolutional layer (note that no residual connections~\cite{he2015deepresiduallearningimage} are used). The time embedding is added to the output of the first of such two convolutional blocks. The encoding blocks are downsampled using a max pooling operation, while before each decoding block an upsampling procedure is performed using transposed convolutional layers. Skip connections between encoding and decoding blocks at each of the 4 levels of the U-Net are added to the output of the decoding blocks. Lastly, the output of the decoder part of the U-Net is processed with a final convolutional block\footnote{This final layer could be made to include a binarization filter as done in Ref.~\cite{Gagnon-Hartman:2021erd}, if the goal is to make this a classification task instead.}. The network is then trained to represent the drift function in Eq.~\eqref{eq:drift_function} by minimizing the loss in Eq.~\eqref{eq:loss}.

The 3D U-Net is implemented using a modified version of the \texttt{unet} package~\cite{perez_garcia_2020_3697931}, which can be found on the \href{https://github.com/NNSSA/Rec21}{GitHub} page of this project.

\subsection{Training and Evaluation}
\label{subsec:training_evaluation}
The dataset comprises 100,000 21-cm lightcones and an equal number of wedge-filtered lightcones, occupying approximately 1.5TB of disk space. The data is randomly split into a 90\%/10\% train/test set. We standardize the lightcones using the mean and standard deviation of the training set to avoid leakage, and do this for the original and wedge-filtered data separately.

We train the model using the voxel-wise mean-squared-error (MSE) loss between the network output and the ground truth, as defined in Eq.~\eqref{eq:loss}. The model is optimized using the Adam optimizer~\cite{kingma2017adammethodstochasticoptimization}, with an initial learning rate of $10^{-3}$ that is halved every two epochs. Due to GPU memory constraints, we use a batch size of 3 lightcones (this is in addition to the model loaded). Although the training and test losses flatten out relatively quickly after a couple of epochs, we train for a total of five epochs while making sure that the model does not overfit by monitoring the test loss for any substantial increases. The model's performance is evaluated by comparing the MSE loss of wedge-recovered lightcones to the identity loss (MSE loss between the original and wedge-filtered data) and by comparing the voxel distribution of the wedge-recovered lightcones to that of the ground truth. Our base model has approximately 43M parameters\footnote{We found that smaller models with around 11M parameters perform reasonably well, but we will use the 43M parameter model for illustrative purposes throughout the paper.} and training takes roughly three days on an 80GB Nvidia A100 GPU. Pre-trained models are made available on the \href{https://github.com/NNSSA/Rec21}{GitHub} page of the project as well.

\subsection{Sample Generation}
\label{subsec:sample_generation}
With the trained model at hand, it now approximates the velocity field along which we can push an initial wedge-filtered box $x_{21}^\mathrm{wf}$ to obtain a sample from the conditional density $\rho(x_{21}|x_{21}^\mathrm{wf})$. We do this by discretizing the time interval $t \in [0,1]$ into 500 steps along which we solve the SDE in Eq.~\eqref{eq:sde}. Specifically, we substitute the trained neural network model for the drift function $\hat{v}$ and solve it using a custom SDE solver based on Heun's second-order method. By repeating this procedure many times, we get a distribution of reconstructed boxes that is effectively marginalized over cosmological initial conditions and the astrophysical parameters that we allowed to vary in the dataset generation. Each sample generation with our 43M parameter model takes approximately 2.5 minutes on an Nvidia A100 GPU, but this time can be reduced to 30 seconds when using a smaller model or fewer time steps.

\section{Results}
\label{sec:results}

\begin{figure}[t!]
    \centering
    \includegraphics[width=\linewidth]{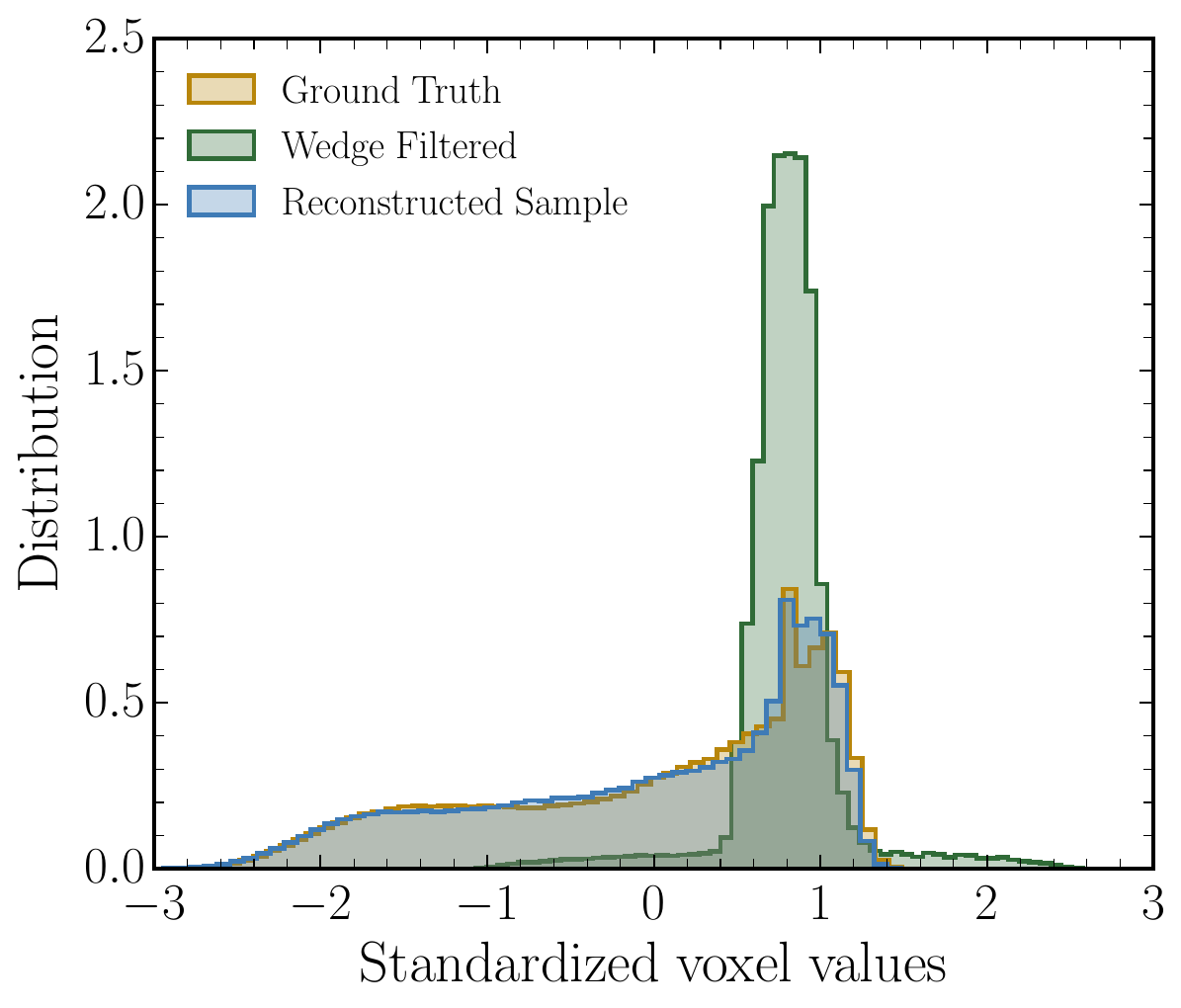}
    \caption{Distribution of voxel values for the ground-truth (yellow), wedge-filtered (green), and reconstructed (blue) lightcones, as illustrated in Fig.~\ref{fig:reconstruction}. The generated samples show a distribution that is much more consistent with that of the ground truth.}
    \label{fig:distributions}
\end{figure}

Having discussed our methodology, we can now evaluate our model's ability in reconstructing 21-cm lightcones. An example of a sample obtained by solving the SDE in Eq.~\eqref{eq:sde} is presented in Fig.~\ref{fig:reconstruction}. The top and bottom rows show slices along the transverse and line-of-sight directions, respectively. The leftmost column displays the wedge-filtered image at $t = 0$. The subsequent columns illustrate the image's evolution as it transforms into a wedge-recovered sample by $t = 1$. This can be compared to the ground-truth image in the sixth column. We repeat this process 2000 times and display the average of these samples in the last column. From this, we observe that large-scale correlations are reasonably well reconstructed, while small-scale features exhibit more randomness (that get averaged out in the last column). This outcome aligns with our discussion in Sec.~\ref{subsec:data_and_model}, where we noted that wedge filtering removes small-scale features more aggressively. We can also examine the resemblance between the ground-truth and reconstructed lightcones by analyzing the voxel distributions of both, see Fig.~\ref{fig:distributions}. This figure highlights the strong similarity between the two lightcones, even at the distribution level.

\begin{figure*}[t!]
    \centering
    \includegraphics[width=0.99\linewidth]{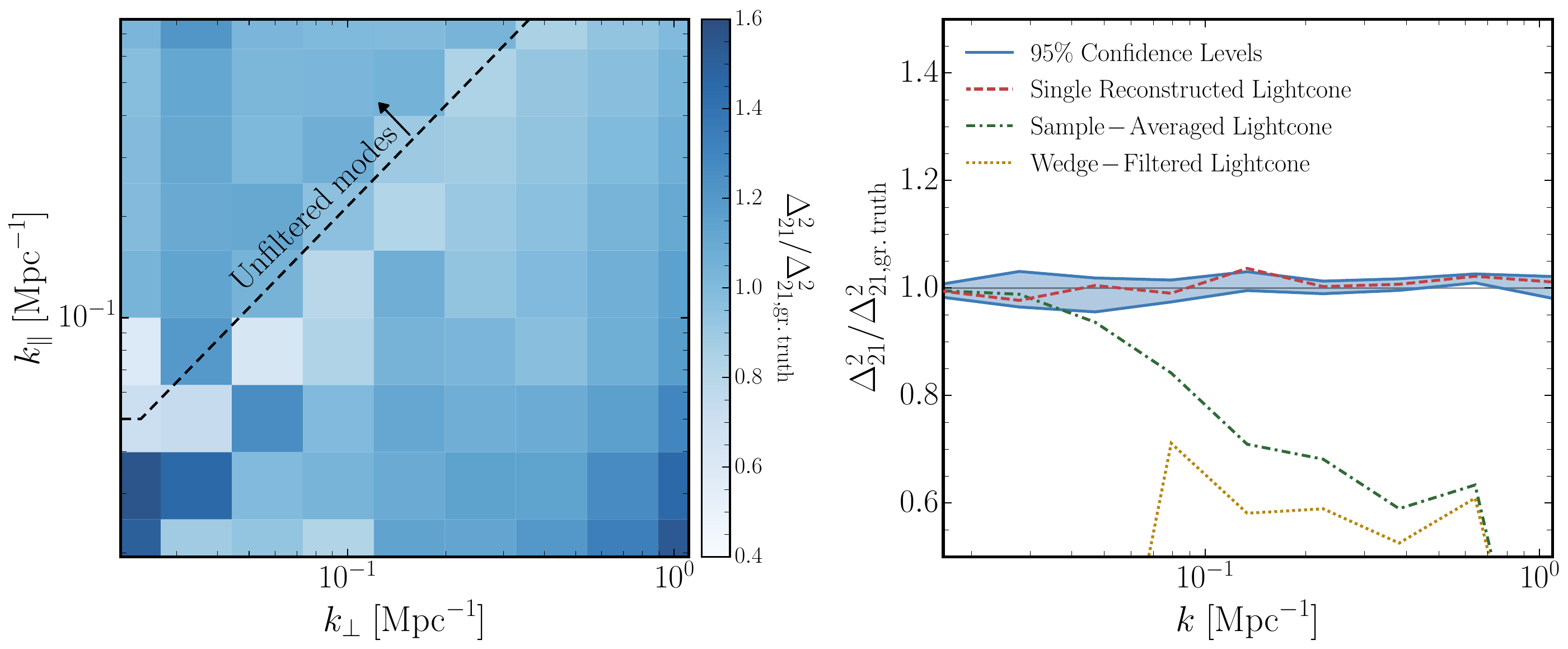}
    \caption{Recovered power spectra of the 21-cm signal. \textbf{Left:} 2D power spectrum in $(k_\parallel, k_\perp)$-space of the wedge-recovered sample in Fig.~\ref{fig:reconstruction} at $t = 1$ relative to the ground truth. The modes to the left of the dashed curve are those initially used by the model to reconstruct the lightcone, while the ones to the right were zeroed out. \textbf{Right:} Spherical power spectra relative to the ground truth. We generate 2000 wedge-recovered lightcone samples, compute their power spectra, and plot the 95\% confidence levels in blue. The red dashed curve shows the power spectrum of a single sample of these ($t=1$ column in Fig.~\ref{fig:reconstruction}), and the green dash-dotted curve is that of the sample-averaged lightcone (rightmost column in Fig.~\ref{fig:reconstruction}). The yellow dotted curve represents the wedge-filtered lightcone.}
    \label{fig:power_spectrum}
\end{figure*}

We further illustrate the reconstruction quality by plotting power spectra in Fig.~\ref{fig:power_spectrum}. The left panel shows the 2D power spectrum for the reconstructed sample in Fig.~\ref{fig:reconstruction} relative to the ground truth. Note that the Fourier modes to the right of the black dashed curve were initially zeroed out, as depicted in the top panel of Fig.~\ref{fig:wedge_plots}. Yet, we can recover the power within tens of percents in that region, even deep inside the foreground wedge. The right panel of Fig.~\ref{fig:power_spectrum} shows the spherical power spectrum and highlights a key feature of this generative model. Namely, we can compute a power spectrum for each of the 2000 generated samples and obtain a distribution as a function of spherical wavenumber $k = (k_\parallel^2 + k_\perp^2)^{1/2}$. This then enables us to set confidence intervals on the retrieved power spectrum (blue solid curves). 
Our reconstructed power spectrum is within approximately 5\% of the ground truth for all $k$ values considered. This is in contrast to the sample-averaged lightcone case (green
dash-dotted curve in this plot) and the wedge-filtered lightcone (yellow dotted curve). 
Our approach not only demonstrates an unbiased recovery of the power spectrum at all $k$ --- even below the foreground wedge --- but also accounts for uncertainties due to unknown cosmological initial conditions and astrophysics. Note that without such a reconstruction, the spherical power spectrum cannot be reliably computed from the wedge-filtered data alone~\cite{Pober:2014lva}.

The MSE loss of the reconstructed lightcones is approximately 51, which can be compared to the identity MSE loss of 655 and the variance of the original lightcone of 429. Overall, the reconstruction shows a high level of accuracy based on visual inspection, distribution-level comparisons, and loss metrics.

\section{Future Avenues}
\label{sec:future}
We have demonstrated the potential of generative modeling techniques in reconstructing 21-cm lightcones distorted by wedge filtering, showing in particular that it is feasible to recover both tomographic maps and power spectra. In this section, we outline several avenues for improving the fidelity and utility of this method, which could strengthen its application in future studies.\\

\noindent
\textbf{Reconstruction within observational conditions.} The dataset used in our training process was directly generated by the \texttt{21cmFAST} simulation code. A more realistic scenario would incorporate instrumental noise and possible systematics from experiments like HERA~\cite{DeBoer:2016tnn} or SKA~\cite{Carilli:2004nx}, see e.g. Refs.~\cite{Gagnon-Hartman:2021erd, Kennedy:2023zos}, where each training input includes a realization of the expected noise. In this way, the learned conditional distribution of wedge-recovered lightcones would be marginalized over variations in experimental conditions as well. 
Note that the typical field as will be observed by SKA or HERA is far larger ($\mathcal{O}(500)\,\mathrm{Gpc}^3$ for SKA~\cite{Santos:2015gra}) than we can directly simulate. Expanding our framework to accommodate larger spatial configurations with different resolutions would be a valuable next step in evaluating its applicability within a practical setting.\\

\noindent
\textbf{Latent-space generative modeling.} The backbone of our model is a 3D U-Net that processes a 3D 21-cm lightcone to produce a corresponding 3D velocity field. In the intermediate steps, each layer maintains its 3D structure, making the process computationally expensive. A recent advancement that could mitigate such computational costs involves latent generative models~\cite{vahdat2021scorebasedgenerativemodelinglatent, rombach2022highresolutionimagesynthesislatent}. For instance, latent diffusion models have been effectively used to generate images and other outputs without any significant loss in quality~\cite{rombach2022highresolutionimagesynthesislatent}. In our context, this approach would involve performing the wedge reconstruction in latent space. Specifically, an encoder would first transform the 3D input box into a 2D or 1D representation, where the stochastic interpolant method could be applied. The resulting solutions would then be converted back to 3D using a decoder. This method would significantly increase the efficiency for both training and sample generation when solving stochastic differential equations.\\

\noindent
\textbf{Scalability.} Another crucial consideration for this method is its performance in higher-dimensional parameter spaces. While in this paper we have varied cosmological initial conditions and three key astrophysical parameters, the 21-cm signal depends on many other parameters that should also be considered~\cite{Mesinger:2010ne, Murray:2020trn, Munoz:2023kkg}. These include both cosmological parameters~\cite{Liu:2015txa}, as well as astrophysical parameters that encapsulate the strength of feedback processes or the spectral distributions of the first galaxies~\cite{Kern:2017ccn}. Consequently, more training data would be required to ensure the model remains effective in the reconstruction process over the entire parameter space. Depending on the availability of computational resources, it may be necessary to use informed priors from other 21-cm analyses, e.g. standard wedge-removed power-spectrum inferences~\cite{Park:2018ljd, Mason:2022obt, Prelogovic:2023uww}, to reduce the need for an exorbitant number of simulation boxes. Another important factor to consider is the model's performance on mock data generated by other simulators~\cite{santos10, visbal12} than the one it was trained on, as this variability can impact the robustness of the results~\cite{Villaescusa-Navarro:2022twv, CAMELS:2023ywa}. \\

\noindent
\textbf{Parameter inference.} The stochastic interpolant framework also enables the inference of likelihoods for the generative model. As outlined in Sec.~\ref{subsec:data_and_model}, our current method uses a dataset where each box is characterized by a unique set of astrophysical parameters which are not directly provided to the network, resulting in a distribution that is marginalized over these parameters. Instead, by conditioning the network on specific parameters of interest (e.g., cosmological or astrophysical), we can retrieve the likelihood of the data. Such an approach has recently been applied within astrophysical and cosmological contexts at the field level with normalizing flows and diffusion models~\cite{Dai:2022dso,Mudur:2022gfq,Legin_2023, Cuesta-Lazaro:2023zuk}. The resulting likelihood can then be used for parameter inference through methods such as simulation-based inference~\cite{thomas2020likelihoodfreeinferenceratioestimation,Leclercq:2021ctr,Hahn:2022wgo, SimBIG:2023ywd,Rose:2024xcb}.\\

\noindent
\textbf{21-cm cosmology.} 
Given the inherent uncertainties in the astrophysics of the first galaxies, standard-ruler analyses are key for robustly extracting cosmological information from high-redshift 21-cm maps. However, the signatures of both the usual baryon acoustic oscillations (BAOs) ~\cite{SDSS:2005xqv} and the high-redshift velocity-induced acoustic oscillations (VAOs)~\cite{Tseliakhovich:2010bj,Munoz:2019fkt} appear at relatively low wavenumbers $k$ (set by the sound horizon $r_\mathrm{s}\approx 150\,\mathrm{Mpc}$), where the foreground wedge is most relevant. The presence of the wedge has so far limited us to computing the 21-cm power spectrum only along the line of sight ($k_\parallel$), allowing for a measurement of the Hubble rate. However, this prevents foreground-removed BAO and VAO analyses from measuring the angular diameter distance (as it is derived from the perpendicular modes $k_\perp$), which carries integrated information from today to high redshifts. A reconstruction of the modes lost due to wedge filtering could allow for BAO and VAO measurements in the $k_\perp$ direction as well, determining the angular diameter distance  and thus nailing down the cosmology of the high-redshift Universe. Although this process might first seem daunting due to the need for numerous simulation boxes, recent advancements in fast semi-analytic simulations for 21-cm astrophysics and cosmology may accelerate this effort~\cite{Munoz:2023kkg, Cruz:2024fsv}.

\section{Conclusions}
\label{sec:conclusion}

In this paper, we introduced a generative model designed to reconstruct 21-cm lightcones that were distorted due to foreground-wedge filtering. Our approach leverages the framework of stochastic interpolants, where a 3D neural network is trained to represent a velocity field that transports wedge-filtered lightcones onto wedge-recovered lightcones. This allows us to construct a generative model capable of sampling from the distribution of wedge-recovered lightcones conditioned on an input wedge-filtered lightcone. These samples are effectively marginalized over unknown cosmological initial conditions and astrophysical parameters.

A significant advantage of a reconstruction approach as presented in this paper is that it eliminates the need for foreground modeling in the wedge region, a notoriously challenging task given that the noise is several orders of magnitude stronger than the 21-cm signal of interest. Fortunately, the inherently non-Gaussian nature of the 21-cm signal enables a scalable model like a neural network to learn correlations between Fourier modes inside and outside of the wedge, allowing a recovery of the lost modes.

We demonstrated the potential of our method by applying it to 21-cm data from simulations. Wedge filtering severely distorts 21-cm lightcones, altering the morphologies of features and complicating their interpretation. The results presented in this work indicate that our model can recover a substantial portion of the lost Fourier modes, producing a reconstructed lightcone that closely resembles the original. We further highlighted the reconstruction quality by computing the 21-cm power spectrum in ($k_\parallel, k_\perp$)-space and showed how we can recover its amplitude within tens of percents accuracy, even deep within the foreground wedge (see Fig.~\ref{fig:power_spectrum}). Moreover, we calculated the spherical power spectrum and set confidence intervals to account for uncertainties in cosmological initial conditions and astrophysical parameters, which will be key for extracting astrophysics and cosmology from upcoming 21-cm observations.

This work builds upon previous studies that explored the application of machine learning tools in 21-cm image analyses. While more investigation is necessary to further assess the applicability of our proposed approach to observational data, generative modeling techniques have already had a significant impact on various research disciplines and could pave the way for new advancements in the 21-cm field.

Our code, pre-trained models, and scripts used to produce the results in this paper can be found at \href{https://github.com/NNSSA/Rec21}{https://github.com/NNSSA/Rec21}.

\section*{acknowledgements}
\noindent We are very thankful to Michael Albergo, Carolina Cuesta-Lazaro, and Adrian Liu for insightful discussions. We also thank the organizers of the `Astro x ML Hackathon 2024' workshop at the Institute for Artificial Intelligence and Fundamental Interactions for their hospitality. NS was supported by a Horizon Fellowship from Johns Hopkins University. RR was supported by National Science Foundation Grant No.~PHY-2210562. JBM was supported at UT Austin by National Science Foundation Grant No.~AST-2307354.
SM is partly supported by the U.S. Department of Energy, Office of Science, Office of High Energy Physics of U.S. Department of Energy under grant Contract Number  DE-SC0012567. 
This work is supported by the National Science Foundation under Cooperative Agreement PHY-2019786 (The NSF AI Institute for Artificial Intelligence and Fundamental Interactions, \href{http://iaifi.org/}{http://iaifi.org/}). TY is funded by the Samsung Science Technology Foundation under Project Number SSTF-BA2201-06. This work utilized the Advanced Research Computing at Hopkins core facility, which is supported by the National Science Foundation grant number OAC1920103.

We acknowledge the use of the following software: \texttt{21cmFAST}~\cite{Mesinger:2010ne, Murray:2020trn}, 
\texttt{unet}~\cite{perez_garcia_2020_3697931},
\texttt{PyTorch}~\cite{paszke2019pytorchimperativestylehighperformance},  \texttt{TorchDyn}~\cite{poli2020torchdynneuraldifferentialequations}, \texttt{torchdiffeq}~\cite{chen2019neuralordinarydifferentialequations}, \texttt{Numpy}~\cite{Harris_2020}, \texttt{SciPy}~\cite{Virtanen_2020}, and \texttt{Matplotlib}~\cite{4160265}.

\bibliography{biblio}
\end{document}